\documentstyle[12pt]{article}
\begin{document}
\large
\date{ }
\begin{center}
{\Large On the experimental search for neutron$\rightarrow$mirror neutron
oscillations}

\vskip 0.7cm

Yu.N.Pokotilovski\footnote{e-mail: pokot@nf.jinr.ru;
tel: 7-49621-62790; fax: 7-49621-65429}

\vskip 0.7cm
            Joint Institute for Nuclear Research\\
              141980 Dubna, Moscow region, Russia\\
\vskip 0.7cm

{\bf Abstract\\}

\begin{minipage}{130mm}
\vskip 0.7cm
Fast neutron$\rightarrow$mirror neutron ($n\rightarrow n^{'}$) oscillations
were proposed recently as the explanation of the GZK puzzle. We discuss
possible laboratory experiments to search for such oscillations and to improve
the present very weak constraints on the value of the
$n\rightarrow n^{'}$ oscillation probability.

\end{minipage}
\end{center}

\vskip 0.3cm


\vskip 0.2cm


\vskip 0.6cm
\section{Introduction.}

Mirror asymmetry \cite{Lee} of our world is a well established fact. The idea
that the Nature, if not P-symmetric, is CP-symmetric \cite{CP} has not been
supported by experiment. But in the seminal paper \cite{Lee}, where the
parity non-conservation hypothesis was first proposed, the existence was
suggested of new particles with the reversed asymmetry: "If such asymmetry is
indeed found, the question still could be raised whether there could not exist
corresponding elementary particles exhibiting opposite asymmetry such that
{\bf in the broader sense there will be over-all right-left symmetry}".
According to \cite{Lee} the transformation in the particle space responsible
for the space inversion is not a simple reflection $P: \vec r\rightarrow
-\vec r$, but a more complicated $PR$ transformation, where $R$ is the
transition of the particle into the reflected state in the mirror particle
space.  From this point of view the Nature is PR-symmetric, the equivalence
between left and right is restored.

This idea has been revived \cite{Kob} after the observation of CP-violation.
In this paper it was shown that mirror particles, if they exist, can not
interact with usual particles through strong or electromagnetic interaction,
but only through some weak and predominantly through gravitational
interactions. They proposed also that mirror particles and massive objects can
be present in our Universe.

Many new ideas appeared during the last forty years on this subject.
There were found strong arguments that the dark matter in the Universe may be
the mirror matter.
Serious implications of the experimental search for the dark matter were
discussed recently. For example the results of DAMA \cite{DA} and
CRESST \cite{CRE} experiments were interpreted \cite{rev2} as the evidence of
scattering of mirror particles in the detectors.
 Mirror matter concept has found also development from superstring
theories.  The recent reviews of the state of art of theoretical and
experimental investigations in the field of mirror particles may be found
in \cite{rev2,rev1}.

The idea was put forward recently \cite{n'pro} (see also \cite{n'resp}) that
fast $n\rightarrow n^{'}$ oscillations could provide a very effective
mechanism for transport of ultra high energy cosmic protons, with the energy
exceeding the Greisen-Zatsepin-Kuzmin cutoff  $5\times 10^{19} eV$, over very
large cosmological distances.

Irrespective of this particular mechanism it turned out that existing
experimental constraints on $n\rightarrow n^{'}$ oscillations are very weak.
The experimental limit on the neutron$\rightarrow$antineutron oscillation
time is strong enough \cite{n-n} due to the high energy release of the
antineutron annihilation $\sim$ 2 GeV. There is no such signal in the case of
$n\rightarrow n^{'}$ transition. Real constraints on the characteristic
time of this process are much smaller than the neutron lifetime \cite{n'pro}.
Indeed, the only signal for $n\rightarrow n^{'}$ transformation is the
disappearance of neutrons from the beam. No special experiment  with the aim
to search for such a disappearance has been performed before. Very rough
estimate of the loss of the neutron beam from the experimental search for the
$n\rightarrow\tilde n$ - oscillations \cite{n-n,n'pro} gives a constraint for
the time of $n\rightarrow n^{'}$ oscillation at the level of 1 s. The neutron
balance in reactors gives not better precision. Since there is no firm
predictions for the probability of the $n\rightarrow n^{'}$ oscillations,
an experimental search for this transition has to be performed with the
highest possible precision.

The present limit on the oscillation time of the $o$--positronium to the
mirror $o$--positronium is $\approx 1$ ms (see experiment \cite{invis} with
reinterpretation in \cite{Gni}). There are plans to improve this limit on
one-two orders of magnitude \cite{Gni}.

The phenomenology of the neutron$\rightarrow$mirror neutron oscillations is
similar to that of neutral kaon, muon$\rightarrow$antimuon and
$n\rightarrow\tilde n$  oscillation.
Starting from $n-n^{'}$ mass matrix
\begin{equation} L=\bar\psi M\psi,
\end{equation}
where spinor
\begin{equation}
\psi={n\choose n^{'}},
\end{equation}
and
\begin{eqnarray}
M=
\left ( \begin{array}{cc} M & \delta m \\
\delta m & M^{'}
\end{array} \right)
\end{eqnarray}
we have standard solution for evolution of the mirror neutron component with
the initial number of ordinary neutrons $n_{0}$:
\begin{equation}
n^{'}(t)=n(0)\frac{\delta m^{2}}{\delta m^{2}+\Delta E^{2}}
sin^{2}\Bigr(\sqrt{\Delta E^{2}+\delta m^{2}}\cdot t\Bigr).
\end{equation}
Here $\delta m$ is the transition mass and $2\Delta E=M-M{'}$ is the mass
difference of the neutron and mirror neutron states.
When oscillations take place in free space the only contribution to $\Delta E$
comes from the neutron interaction with external magnetic field $B$:
$2\Delta E=\mu B$, where $\mu=6\cdot 10^{-12}$ eV/G is the neutron magnetic
moment. Introducing $\tau_{osc}=\hbar/\delta m$ and $\omega=\Delta E/\hbar$ we
obtain
\begin{equation}
n^{'}(t)=\frac{n(0)}{1+(\omega\tau_{osc})^{2}}
sin^{2}\Bigr(\sqrt{1+(\omega\tau_{osc})^{2}}\cdot t/\tau_{osc}\Bigr),
\end{equation}
$\omega\approx 4.8\times 10^{3}$ s$^{-1}$ in the field $B=1$ G.

Since experimentally we have always $\omega\tau_{osc}\gg 1$, two limiting
cases are possible: $\omega t\gg 1$ and $\omega t\ll 1$. In the first case the
average of oscillating term is equal to 1/2, and
\begin{equation}
n^{'}(t)=\frac{1}{2(\omega\tau_{osc})^{2}}.
\end{equation}
The second case gives
\begin{equation}
n^{'}(t)=(t/\tau_{osc})^{2}.
\end{equation}
The second, more experimentally sensitive situation, is realized when coherent
evolution of the wave function $\psi$ takes place in the well magnetically
shielded conditions (from external and the Earth magnetic fields).

Now let us consider possible experimental approaches to the search of
$n\rightarrow n^{'}$ oscillations. There are two such approaches: the neutron
beam experiments and the storage of ultracold neutrons \cite{ucn}.

Two kinds of the beam experiments are possible: the first one -- based on the
measurement of disappearance of neutrons from the beam due to $n\rightarrow
n^{'}$ transformation and the second one, when after such hypothetical
transformation the incident neutron beam is stopped by the neutron absorber,
and the mirror component then again can be re-transformed to the ordinary
neutron component according to Eq. (6). Let us estimate possible sensitivity.

\section{Disappearance of the neutrons from the beam.}

Let the neutron beam with the flux $\phi_{0}$ and the average velocity
$v$ enters the magnetically shielded neutron flight path of the length $L$.
The flux of mirror neutrons at the end of the flight path is
$\phi_{n{'}}(t)=\phi_{0}(L/v\tau_{osc})^{2}$.
It is just the number of neutrons disappeared from the beam.
To forbid the $n\rightarrow n^{'}$ transformation the magnetic field $B$ such
that $\omega_{B}t\gg 1$ should be switched on along the flight path.
Since the change in counts due to $n\rightarrow n^{'}$ transformation  is
expected to be small, in the limit of one standard error during the time
$T_{exp}$ for each of the measurements -- with permitting and forbidding of
oscillations, we get
\begin{equation}
\phi_{0}(\frac{L}{v\tau_{osc}})^{2}T_{exp}<(2\phi_{0}T_{exp})^{1/2},
\end{equation}
and
\begin{equation}
\tau_{osc}>\frac{L}{v}(\phi_{0}T_{exp}/2)^{1/4}.
\end{equation}
With $\phi_{0}\approx 3\times 10^{7}$ s$^{-1}$\cite{VCN}, $v\approx 100$ m/s,
$L=$5 m, and the experimental time $T_{exp}$=1 month
$\approx 2.5\cdot 10^{6}$ s, we get $\tau_{osc}>$125 s.

\section{Process $n\rightarrow n^{'}\rightarrow n$.}

In this approach the flight path consists of two magnetically shielded
sections with the length of $L/2$ each, with the perfect absorber of
neutrons in the middle. In the first section the neutrons transform to
the mirror state with the probability $w=(L/2v\tau_{osc})^{2}$, then the
incident neutrons are absorbed, and, in the second section the transformation
$n^{'}\rightarrow n$ should take place with the same probability.
The neutron intensity at the end of the flight path is
\begin{equation}
\phi_{n{'}}(t)=\phi_{0}\Bigl(\frac{L}{2v\tau_{osc}}\Bigr)^{4},
\end{equation}
The magnetic field in any of the sections will forbid the oscillations.
If the neutron detector count rate with stopped beam is $\phi_{bgr}$ the same
considerations give:
\begin{equation}
\phi_{0}(\frac{L}{2v\tau_{osc}})^{4}T_{exp}<(2\phi_{bgr}T_{exp})^{1/2},
\end{equation}
with the result
\begin{equation}
\tau_{osc}>\frac{L}{2v}(\phi_{0})^{1/4}
\bigl(\frac{T_{exp}}{2\phi_{bgr}}\bigr)^{1/8}.
\end{equation}
With the same parameters of the experiment and assuming
$\phi_{bgr}=0.01$ s$^{-1}$, we get $\tau_{osc}>20$ s.

\section{The storage of ultracold neutrons.}

The above calculation is not applicable to the ultracold (UCN) storage
experiments \cite{ucn}, where the neutrons are confined in the closed
chambers.  In this case the neutron-wall collisions at the rate $f\approx
<v>/<d>$, where $v$ is the neutron velocity and $d$ is the distance between
collisions, cause decoherence, disrupting the oscillation, the mirror
component being lost, penetrating into the wall with the rate $\lambda\approx
1/f\tau_{osc}^{2}$ in the case of degaussed storage chamber, and with the rate
$\lambda\approx f/2(\omega_{B}\tau)^{2}$ in the magnetic field $B$,
($\omega_{B}=\omega\cdot B (G)$), when the transition to the mirror state is
suppressed. If the neutron lifetime storage measurements are performed in
degaussed and non-degaussed conditions with the precision 1 s, what
corresponds to the uncertainty of $\alpha\sim 10^{-6}$ s$^{-1}$ in the decay
constant, we get the precision for the oscillation time
$\tau_{osc}>1/(f\alpha)^{1/2}$. At the typical $f\sim (5-10)$ we get
$\tau_{osc}>300-500$ s.

No previous neutron lifetime measurements in UCN storage mode were performed
in degaussed conditions. The interesting observation exists, however, that the
measurements of the neutron lifetime by two different methods: the first one
-- the measurement of the neutron density in the decay volume and counting the
$\beta$-decay products (electrons or protons), and the second one -- the
storage of ultracold neutrons, give slightly different results. The table
contains the results of all the measurements, which display the errors not
exceeding 10 s. The first method is insensitive to any invisible decay or
disappearance of neutrons from the decay volume, the second method, to the
contrary, is sensitive. The difference in the results of these methods, if
correct, gives a hint at some invisible channel of disappearance of neutrons
from storage chambers.  Without the result \cite{nlife} the difference in
decay constants is $(5.47\pm 2.85)\cdot 10^{-6}$ s$^{-1}$, with taking into
account \cite{nlife} this difference is $(9.7\pm 2.8)\cdot 10^{-6}$ s$^{-1}$.
The predicted neutron decay constant into a hydrogen atom:  $n\rightarrow
H+\tilde\nu_{e}$ \cite{atom} is as low as $\sim 4\cdot 10^{-9}$ s$^{-1}$,
(branching ratio $\sim 3.8\cdot 10^{-6}$) and can not explain the observed
difference (see also \cite{hydr} where the estimate of the neutron decay
constant to the atomic mode has been performed, based on the earlier neutron
lifetime measurements).

The UCN storage measurements were performed in the Earth magnetic field $\sim
0.5$ G, and using the above expression for the rate of the neutron loss
from the chamber with the neutron-wall collision frequency $f$,
$\lambda=f/2(\omega_{B}\tau_{osc})^{2}$, we get the oscillation time
$\tau_{osc}=(f/2\lambda)^{1/2}/\omega_{B}$.
With $f\approx (5-10)$ s$^{-1}$, $\omega_{B}\sim 2.4\cdot 10^{3}$ s$^{-1}$,
and $\lambda=5\cdot 10^{-6}$ s$^{-1}$, we get $\tau_{osc}=0.3-0.4$ s - the
figure close to the necessary one for the mechanism \cite{n'pro}! It is
difficult to say at this moment how seriously this should be taken. But
it is clear that high precision neutron lifetime experiments of both classes
are important.  On the other hand, we should not ignore the obvious trend of
beam experiment lifetime data to the lesser values, in the direction of better
agreement with the neutron storage results.

\section{UCN flow experiment.}

This approach seems somewhat simpler than the precise UCN storage
measurements. Consider UCN constant flow through the storage chamber with
entrance and exit windows (it is assumed isotropic angular distribution). The
neutron balance equation has the view:
\begin{equation}
\frac{dN}{dt}=\phi_{0}s_{in}-\rho\frac{Sv}{4}\mu-
\rho\frac{(s_{in}+s_{out})v}{4}-\frac{N}{\tau_{n}}=0.
\end{equation}
Here $N$ is the equilibrium number of neutrons in the chamber, $\phi_{0}$ is
the neutron flux density at the entrance to the chamber, $s_{in}$ and
$s_{out}$ are the areas of entrance and exit windows, $\rho$ is the
neutron density in the chamber, $v$ is the mean neutron velocity, $V$ and $S$
are the volume and the area of the internal surface of the chamber, $\mu$ is
the neutron loss probability per bounce, $\tau_{n}$ is the neutron
$\beta$-decay lifetime. The first term at the right side is the neutron influx
to the chamber, the second one is the neutron loss due to collisions with the
internal surface, the third one is the neutron efflux to both holes, and the
last one is the neutron $\beta$-decay.

If $\phi_{0}=const$, the equilibrium neutron density
\begin{equation}
\rho=\frac{4\phi_{0}s_{in}}{v(S\mu+s_{in}+s_{out}+\delta)},
\end{equation}
where $\delta=4V/v\tau_{n}$.

We can estimate the sensitivity of equilibrium neutron density to the change
of neutron loss coefficient
\begin{equation}
\frac{d\rho}{d\mu}=-\frac{4\phi_{0}Ss_{in}}{v(S\mu+s_{in}+s_{out}
+\delta)^{2}}.
\end{equation}
The detector count rate is
\begin{equation}
I_{det}=\rho s_{out}v/4=
\frac{\phi_{0}s_{in}s_{out}v}{S\mu+s_{in}+s_{out}+\delta}.
\end{equation}
Its variations because of variation $\Delta\mu$ of the neutron loss
probability is
\begin{equation}
\Delta I=-\frac{\phi_{0}Ss_{in}s_{out}\Delta\mu}
{\biggl(S\mu+s_{in}+s_{out}+\delta)^{2}}.
\end{equation}

At the effect of oscillations in the limit of one error during
the measurement times $T_{exp}$ with and without the magnetic field we have:
\begin{equation}
\frac{1}{(f\tau_{osc})^{2}}
\frac{\phi_{0}Ss_{in}s_{out}T_{exp}}
{(S\mu+s_{in}+s_{out}+\delta)^{2}}<
\Biggl(\frac{2\phi_{0}Ss_{in}s_{out}T_{exp}}
{S\mu+s_{in}+s_{out}+\delta}\Biggr)^{1/2}.
\end{equation}

Finally, we have
\begin{equation}
\tau_{osc}>\frac{1}{2^{1/4}f}\frac{(T_{exp}\phi_{0}s_{in}s_{out})^{1/4}
S^{1/2}}{(S\mu+s_{in}+s_{out}+\delta)^{3/4}}.
\end{equation}
At $\mu\sim 10^{-5}$ achieved with the Fomblin oil or grease cover of the
wall surface (see Fig. 1 in the recent review \cite{Pok}), reasonably
$S\mu\ll s_{in},s_{out}$. With the $f\approx <v>/<d>$, $<d>\sim 4V/S$, 
taking $s_{in}=s_{out}=s$, demanding that the neutron decay does not decrease 
essentially the neutron density in the chamber:  $s\gg 4V/(\tau_{n}v)$, and 
assuming that the storage chamber is a cylinder with the diameter and the 
height $l$ we get:
\begin{equation}
\tau_{osc}\approx 0.13\biggl(\frac{T_{exp}\phi_{0}}{s}\biggr)^{1/4}
\frac{l^{2}}{v}.
\end{equation}

During one month measurements with and without the magnetic field
($T_{exp}\approx 2.5\cdot 10^{6}$ s), with
$\phi_{0}=10^{3}$ cm$^{-2}$ s$^{-1}$, $l$=100 cm, $s$=10 cm$^{2}$, and
$v$=400 cm/s, we get $\tau_{osc}\approx$ 420 s.

We neglected the gravity in this estimate. More detailed optimization of the
experiment in respect to all parameters taking into account the effect
of gravity may increase the sensitivity considerably.

In all described types of experiments the switching on weak magnetic field to
forbid $n\rightarrow n^{'}$ oscillations, and increasing the detector's count
rate, can, in principle, produce the opposite effect due to neutron
reflection from small magnetic potential $U=\mu B$. For the smooth
potential this effect is exponentially small -- the reflection
coefficient for the neutrons with the energy $E$ incident at the potential
of the height $U_{0}\ll E$, which is changing continuously over the region
$x_{0}$, is proportional to $\sim exp(-4\pi kx_{0})$, where $k$ is the neutron
wave vector. In the worst case $x_{0}\sim 10$ cm, $k\sim 10^{6}$ cm$^{-1}$.

\newpage
\begin{center}
The results of the neutron lifetime measurements in the beam experiments and
in the UCN storage experiments.  Only the results with uncertainties not
exceeding 10 s were taken into consideration.
\end{center}

\begin{tabbing}
Beam experimentsqqqqqqqqqqqqq\=qqqqqqqqqqqqqqqqqqqqq\=\kill
{\bf Beam experiments}            \> {\bf Storage experiments} \\
								\\
891$\pm$9 (1988)\cite{Spi}               \> 877$\pm$10 (1989)\cite{Paul} \\
893.6$\pm$3.8$\pm$3.7 (1990)\cite{Byr1}  \> 870$\pm$8 (1989)\cite{Khar} \\
889.2$\pm$3.0$\pm$3.8 (1996)\cite{Byr2}  \> 887.6$\pm$3.0 (1989)\cite{Mam}\\
886.8$\pm$1.2$\pm$3.2 (2003)\cite{Dew}   \> 888.4$\pm$3.3 (1992)\cite{Nesv}\\
                                  \> 882.6$\pm$2.7 (1993)\cite{Mor} \\
                                  \> 885.4$\pm$0.9$\pm$0.4 (2000)\cite{Arz}\\
                                  \> 881.$\pm$3.0 (2000)\cite{Mam2} \\
                               \> 878.5$\pm$0.7$\pm$0.3 (2004)\cite{nlife} \\
{\bf Averaged value}          \>{\bf Average value without\cite{nlife}}\\
 889.2$\pm$2.4                    \>  884.9$\pm$0.8         \\
         		 \>{\bf Average value including\cite{nlife}} \\
				   \> 881.6$\pm$0.6
\end{tabbing}

\end{document}